\documentclass[twoside]{article}
\usepackage{fleqn,espcrc2}
\usepackage{graphicx}

\title{\vbox to 0pt{\vskip-1in
\rightline{\normalsize Preprint MPI-PhT/2000-03}}\vskip-8pt
Dark Matter at the Galactic Center}

\author{Paolo Gondolo \address{Max-Planck-Institut f\"{u}r Physik,
    F\"{o}hringer Ring 6,
    80805 M\"{u}nchen, Germany (gondolo@mppmu.mpg.de)} 
and Joseph Silk
  \address{Astrophysics, University of Oxford, Keble Road,
    Oxford, OX1 3RH, U.K. \& Department of Astronomy and Physics,
    University of California, Berkeley, CA 94720
    (silk@astro.ox.ac.uk)}
}

\begin{document}

\begin{abstract}
Particle dark matter near the galactic center is accreted by the central black
hole into a dense spike, strongly enhancing its annihilation rate.  Searching
for its annihilation products may give us information on the presence or
absence of a central cusp in the dark halo profile.
\vspace{1pc}
\end{abstract}

\maketitle

This is a summary of a paper of ours, ref.~\cite{gs99}, in which we
use the absence of neutrino signals from the galactic center to bound the
steepness of a possible central cusp in a dark matter halo made of neutralinos.
This summary updates the bounds published in~\cite{gs99} by using the
current upper limit by the MACRO collaboration~\cite{macro} on the neutrino
emission from the galactic center.\footnote{Thanks to Francesco Ronga for
  communicating the new upper limit.}

The evidence is mounting for a massive black hole at the galactic center. Ghez
et al.~\cite{ghe98} have confirmed and sharpened the Keplerian behavior of the
star velocity dispersion in the inner 0.1 pc of the galaxy found by Eckart and
Genzel~\cite{eck9697}. These groups estimate the mass of the black hole to be $
M = 2.6 \pm 0.2 \times 10^6 \, M_{\odot}$.

If cold dark matter is present at the galactic center, as in current models of
the dark halo, it is accreted by the central black hole into a dense spike.
Particle dark matter then annihilates strongly inside the spike, making it a
compact source of photons, electrons, positrons, protons, antiprotons, and
neutrinos. 

The spike luminosity depends on the density profile of the inner
halo: halos with finite cores have unnoticeable spikes, while halos with inner
cusps may have spikes so bright that the absence of a detected neutrino signal
from the galactic center already places interesting upper limits on the density
slope of the inner halo. 

Figure 1 illustrates the two classes of halo models. The ``empirical'' models
are fit to the data and have a central region of constant density, called the
core.  The ``theoretical'' models arise from the results of numerical N-body
simulations and have a power law density profile in the inner region, dubbed
the cusp.  Actually, even the highest resolution results presently available
extend inwards only to $\sim 1$ kpc, but we have boldly extrapolated the cusp
to the inner parsec.

The effect of the central black hole on the dark matter distribution in its
neighborhood can be found in the following way.  Before the formation of the
black hole, the dark matter density within its radius of influence ($\sim 0.2$
pc) can be assumed to be either constant or a power law of index $\gamma$
($\rho \sim r^{-\gamma}$). The dark matter density after the formation of the
black hole is obtained by assuming that the black hole grows slowly and hence
the dark matter distribution evolves adiabatically. Conservation of the three
adiabatic invariants -- phase-space density, angular momentum, and radial
action -- then gives the final dark matter density.

A power-law density profile results around the black hole, with an index
$\gamma_{\rm spike}$ that depends on the initial index $\gamma$ and on the
analytical properties of the initial profile, i.e. core or cusp. We call
``spike'' this density enhancement close to the black hole, to distinguish it
from the cusp further out. The maximum density in the spike is reached either 
at the small distance of $\sim 10$ Schwarzschild radii within which dark
matter is captured by the black hole, or at a larger radius where the
annihilation time becomes equal to the age of the black hole and within which
the density is constant. Examples of spike density profiles are given
in~\cite{gs99}. 

\begin{figure}[t]
\label{fig:profile}
\includegraphics*[width=7.5cm]{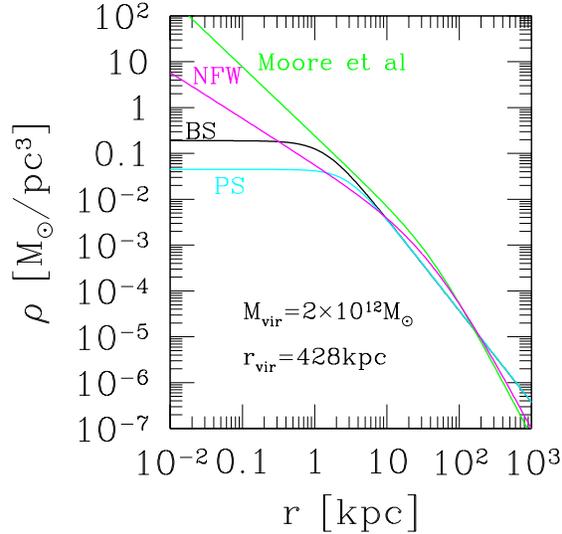}
\vspace{-3\baselineskip}
\caption{Two classes of dark matter profiles for our galactic halo: with a
  central core (BS~\protect\cite{bs} and PS~\protect\cite{ps}) and with a
  central cusp (NFW~\protect\cite{nfw} and Moore et al.~\protect\cite{moore}).}
\end{figure}

Annihilation signals, which increase with the square of the density, are
enhanced dramatically. The enhancement increases with increasing initial slope
$\gamma$, and this allows an upper limit to be set on the value of $\gamma$
given an upper limit on some of the annihilation signals from the galactic
center. In~\cite{gs99} we considered the neutrino emission. High energy
neutrinos from the galactic center could be detected with a neutrino telescope
in the Northern hemisphere through their conversion to muons in a charge
current interaction in the rock surrounding the detector. 

\begin{figure}[t]
\label{figgmaxnu}
\vspace{-3.7\baselineskip}
\includegraphics[width=7.5cm]{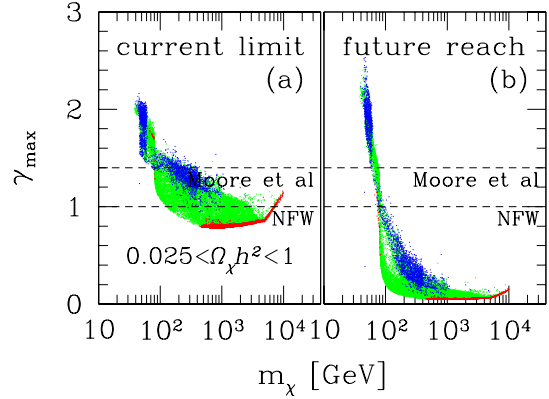}
\vspace{-3.3\baselineskip}
\caption{Maximum inner slope $\gamma$ of the dark matter halo compatible with 
  the upper limit on the neutrino emission from the galactic center. (a)
  Current limit at 1 GeV; (b) future reach at 25 GeV.}
\end{figure}

The current bound~\cite{macro} on the neutrino emission from the galactic
center is 1104 neutrino-induced muons $>1$ GeV per km$^2$ per year. We impose
this bound on the emission expected from neutralino dark matter in the minimal
supersymmetric model, calculated using the DarkSUSY code~\cite{gon99}. We use
the database of points in supersymmetric parameter space built in
refs.~\cite{ber96,eds97,ber98}, namely the 35121 points in which the neutralino
is a good cold dark matter candidate, in the sense that its relic density
satisfies $0.025 < \Omega_\chi h^2 < 1 $.  The upper limit comes from the age
of the Universe, the lower one from requiring that neutralinos are a major
fraction of galactic dark halos.

For each point in parameter space, we can then obtain a separate upper bound
$\gamma_{\rm max}$ on the inner halo slope. These bounds are plotted in
figure~2a.  (Plotted values of $\gamma_{\rm max}>2$ are unphysical
extrapolations but are shown for completeness.) Present bounds are of the order
of $\gamma_{\rm max} \sim 0.5$, right in the ballpark of current results from
N-body calculations.

Future neutrino telescopes observing the galactic center could probe the inner
structure of the dark halo, or indirectly find the nature of dark matter.  For
example, with a muon energy threshold of 25 GeV, the neutrino flux from the
spike after imposing the current constraints could still be over 2 orders of
magnitude above the atmospheric background (Fig.~3), allowing to probe $\gamma$
as low as 0.05 (Fig.~2b).

In conclusion, we have shown that if the galactic dark halo is cusped, as
favored in recent N-body simulations of galaxy formation, a bright dark matter
spike would form around the black hole at the galactic center. A search of a 
neutrino signal from the spike could either set upper bounds on the
density slope of the inner halo or clarify the nature of dark matter. 

\begin{figure}[t]
\label{figphimumax}
\vspace{-4\baselineskip}
\includegraphics[width=7.5cm]{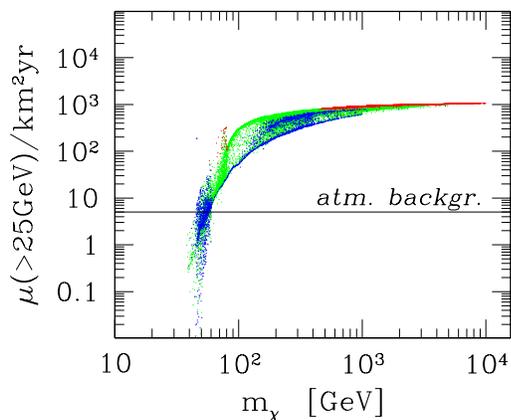}
\vspace{-3\baselineskip}
\caption{Maximal flux of neutrino-induced muons in a neutrino telescope from
  neutralino annihilations at the galactic center, after imposing the current
  constraints on the neutrino emission.}
\end{figure}

\end{document}